\documentclass[journal=jpccck,manuscript=article,layout=twocolumn]{achemso}

\usepackage[utf8]{inputenc}
\usepackage[T1]{fontenc} 
\usepackage{amsmath}
\usepackage{amssymb}

\author{Klaus H. Eckstein\textsuperscript{a}}
\affiliation[uniwue]{\textsuperscript{a}Institute of Physical and Theoretical Chemistry, Julius-Maximilian University W\"urzburg, 97074 W\"urzburg, Germany}
\author{Tobias Hertel\textsuperscript{a}}
\affiliation[uniwue]{\textsuperscript{a}Institute of Physical and Theoretical Chemistry, Julius-Maximilian University W\"urzburg, 97074 W\"urzburg, Germany}
\email{tobias.hertel@uni-wuerzburg.de}
\phone{+49 931 3186300}

\title{Electronic Structure and Scaling of Coulomb Defects in Carbon Nanotubes from Modified H\"{u}ckel Calculations}


\begin{document}

\begin{abstract}
Controlled doping and understanding its underlying microscopic mechanisms is crucial for advancement of nanoscale electronic technologies, especially in semiconducting single-wall carbon nanotubes (s-SWNTs), where adsorbed counterions are known to govern redox-doping levels. However, modeling the associated 'Coulomb defects' is challenging due to the need for large-scale simulations at low doping levels. Using modified Hückel calculations on 120 nm  long s-SWNTs with adsorbed $\rm Cl^-$ ions, we study the scaling properties of shallow Coulomb defect states at the valence band edge and quantum well (QW) states in the conduction band. Interestingly, the QW states may underlie observed exciton band shifts of inhomogeneously doped semiconductors. Binding energies of Coulomb defects are found to scale with counterion distance, effective band mass, relative permittivity and counterion charge according to $d^{\alpha-2}m^{\alpha-1}\epsilon_r^{-\alpha}|z_j|^{\alpha}$, with $\alpha$ as an empirical parameter, deepening our understanding of s-SWNT doping.
\end{abstract}

\section{Introduction}
The unique properties of intrinsic and doped semiconducting single-wall carbon nanotubes (s-SWNTs) have garnered attention for a wide array of applications. These include electronics~\cite{Bishop2020}, sensing~\cite{Chen2016}, photonics~\cite{Avouris2008, He2018, Ishii2018}, photovoltaics~\cite{Ren2011, Kubie2018, Jain2012}, and bioimaging~\cite{Pan2017}. To enhance their functional versatility for such uses, it is critical to be able to control the s-SWNTs' electronic structure through doping.

This can be achieved through charge-transfer- and redox-active species, or by substitutional incorporation. Commonly used oxidizing agents include nitric acid ($\rm HNO_3$)~\cite{Hu2003}, F4TCNQ~\cite{Nosho2007}, viologen~\cite{Kim2009}, triethyloxonium hexachloroantimonate (OA)~\cite{Chandra2010} and others. On the reducing side, alkali metals~\cite{Rao1997} and hydrazine~\cite{Klinke2005} were shown to be effective.

Among oxidizing agents, gold(III) chloride ($\rm AuCl_3$) has been particularly well studied for its effective p-doping of s-SWNTs~\cite{Kim2008, Duong2010, Lee2010, Kim2011, Murat2014, Hertel2019}. This can be attributed to the high redox potentials of gold cations. Notably, the level of p-doping directly correlates with the concentration of adsorbed $\rm Cl^-$ counterions~\cite{Kim2011}, suggesting their significant role for electronic and optical properties of doped s-SWNTs~\cite{Mouri2013, Eckstein2017, Eckstein2019, Eckstein2021}.

High-level quantum chemical calculations for $\rm AuCl_3$-assisted doping have primarily focused on short nanotube unit cells, which are representative of heavy doping~\cite{Duong2010,Murat2014}. However, the treatment of low to moderate doping levels presents challenges to more refined quantum chemical methods due to the large system size involved. To circumvent such limitations, simpler semi-empirical models have been used to explore the formation of what can be described as 'Coulomb defects' - shallow charge traps of a few nanometers in size - resulting from the interaction between charges on the s-SWNT and exohedral counterions~\cite{Eckstein2017}.

In order to provide a more rigorous description of such Coulomb defects at low doping levels, we here use a semi-empirical modified H\"{u}ckel method with a minimal basis set and periodic boundary conditions. Our approach includes curvature corrections to resonance integrals and adjustments to Coulomb integrals to account for the interaction of $\pi$ electrons with exohedral counterions.

The modified H\"{u}ckel calculations effectively capture the energetics and wavefunctions of Coulomb defects, along with their dependence on both internal and external parameters. When compared to a variational analysis that examines the scaling properties of these defects, we can accurately describe how they are influenced by factors such as counterion charge, distance, effective band mass, nanotube chirality, and relative permittivity. In addition, we uncover quantum well-like behavior in conduction band states between counterions, suggesting potential implications for excitonic states.

\section{Methods}

In this study, we employ a modified H\"{u}ckel model, focusing solely on the $p_z$ orbitals of carbon atoms to represent the $\pi$-system of the nanotube as a minimal basis set. The same minimal basis set without curvature corrections has previously also been used to approximate quasiparticle energies - specifically when describing the scaling properties of excitonic features in s-SWNTs \cite{Perebeinos2004}.

To account for electrostatic interactions between $\pi$ electrons and exohedral counterions, we introduce a correction to the Coulomb integrals, $\alpha_i$. Specifically, these interactions are represented by $e^2\, z_j/(4\pi\epsilon_0\epsilon_r |r_i-r_j|)$ where $z_j$ is the charge number of the counterions located at positions $r_j$.

Dielectric screening is accounted for by using a relative permittivity $\epsilon_r$ of 4, in line with previous work \cite{Perebeinos2004}. The effect of screening on the electronic structure is further elaborated in the main discussion. 
\begin{figure}[htbp]
	\centering
		\includegraphics[width=8.4 cm]{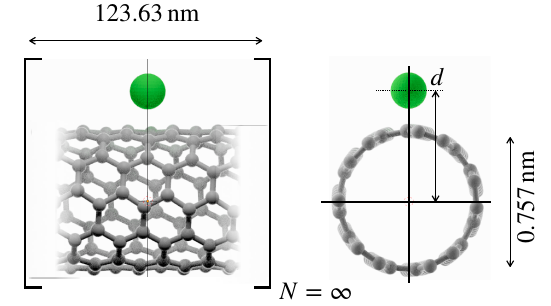}
		\caption{{\bf Nanotube-Counterion Geometry:} Geometry for modified H\"{u}ckel calculations, with counterion distance $d$. Tube segment contains 10,920 C atoms per 123.63 nm.}
		\label{fig1}
\end{figure}

Additionally, we apply corrections to the nearest-neighbor resonance integrals, $\beta$, commonly designated as $\gamma_0$ in tight-binding nomenclature, to account for curvature-induced modifications \cite{Ding2002,Hagen2003}. This yields three bond-specific resonance integrals contingent on the bond orientation relative to the circumferential tube direction. The three resonance integrals are determined using the scalar product of reciprocal space unit vector $\bf K_1$ and the bond orientation vector ${\bf r}_i$ ($i=1,2,3$). Following the discussion by Ding et al.~\cite{Ding2002}, we used $\gamma_i=\gamma_0\, \cos(\alpha_i)$, where $\alpha_i=\tfrac{1}{2}\, {\bf K}_1\cdot {\bf r}_i$.

For best agreement between the computed and empirical band gaps of (6,5) s-SWNTs of around 1.6 eV,\cite{Eckstein2017} or similarly large predicted free carrier absorption thresholds in s-SWNTs \cite{Pedersen2004, Perebeinos2004}, we chose a nearest neighbor resonance integral $\gamma_0$ of 4 eV. Although values in the range of 2.3 to 3.0 eV are more commonly used when discussing electronic transport studies, the choice of $\gamma_0$ is not as critical here, since counterion-induced band changes are mainly governed by the aforementioned Coulomb term for electron-counterion interactions. 

The effect of changes to $\gamma_0$ primarily lies in its reciprocal relationship with the effective band mass $m$. This will be addressed later on in our discussion of the variational calculations by including the role of the effective band mass for the kinetic energy term when determining defect binding energies. Accordingly, reducing the resonance integral from 4 to 3 eV only results in a modest 8\% change in the defect state binding energy.

The geometric structure of a (6,5) nanotube, used in our calculations, is illustrated schematically in Fig.~\ref{fig1}. The tube has a diameter of 757 pm, with 364 carbon atoms per unit cell. To emulate doping-induced changes at carrier concentrations in the $0.01\,{\rm nm}^{-1}$ range, we aggregated 30 such unit cells, yielding a total of 10,920 carbon atoms on a 123.56 nm long nanotube segment. Periodic boundary conditions were considered when populating the H\"{u}ckel matrix for this structure.

As reference point for gold(III) chloride doped s-SWNTs, the $\rm Cl^-$ ion distance from the nanotube axis $d=r_{\rm tube}+r_{\rm vdW|C}+r_{\rm vdW|Cl^-}$ is given by a van der Waals distance of 325 pm between carbon atoms and the $\rm Cl^{-}$ ion at a nanotube radius of 379\,pm. This corresponds to a distance of the ion from the nanotube axis of $d=704\, \rm pm$. Unless mentioned otherwise, this as well as $\epsilon_r=4$, $z_j=-1$ and $(n,m)=(6,5)$ are reference parameters used in the calculations.

The computational code was written using the matrix algebra functionality of the IGOR software package (Wavemetrics Inc.). Results were validated by comparison with curvature corrected zone-folding band structure calculations within the tight-binding approximation \cite{Ding2002, Hagen2003}.

The manuscript text was developed and edited with the assistance of OpenAI's Chat GPT-4 code interpreter version, accessed via its browser-based command-line interface. The primary objective was to solicit the LLM's assistance in proofreading and iteratively refining human-generated text. A typical prompt to the LLM might read, '\textit{Please proofread, check for clarity, flow, and redundancies: MANUSCRIPT TEXT}'. Subsequently, the texts were fine-tuned using the 'DeepL SE Write' (beta) command line interface (deepl.com) to help preserve the original intent and ideas of their human authors.

\section{Results and Discussion}

\subsection{Scaling of Coulomb Defects}
H\"{u}ckel theory, known for its simplicity and ability to make semi-quantitative predictions about the electronic structure and optical properties of conjugated hydrocarbon systems~\cite{Huckel1931, Hoffmann1963}, has successfully been applied to $\pi$ electron systems such as graphene and carbon nanotubes~\cite{Saito1992, Hamada1992}. It uses a simplified version of the Schr\"{o}dinger equation, considering only a limited number of atomic orbitals while neglecting electron-electron interactions. Given the valuable insights that it offers into the electronic and optical properties of such systems, we have here applied it to study the interaction of a (6,5) and selected (n,0) nanotubes with exohedral charges.

\begin{figure}[htbp]
	\centering
		\includegraphics[width=8.4 cm]{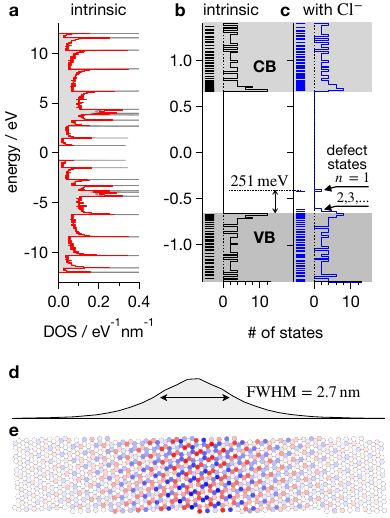}
        \caption{{\bf H\"{u}ckel Calculations for (6,5) s-SWNT.} a) DOS comparison: modified H\"{u}ckel (red) vs. zone-folding. b) Band gap DOS detail. c) DOS with a single $\rm Cl^-$, showing defect states at the VB edge. d) $n=1$ defect state charge density. e) $\pi$ orbital sign and amplitude in false color representation.}
		\label{fig2}
\end{figure}

In Fig.~\ref{fig2}a), the density of states (DOS) for a (6,5) nanotube is represented by the red histogram. The grey shaded DOS is derived from the tight-binding graphene band structure~\cite{Wallace1947} including curvature corrections,\cite{Ding2002,Hagen2003} within the well-established zone-folding scheme~\cite{Saito1992}. The excellent overlap between the two DOS graphs validates the formal equivalence of these methods.

Figure \ref{fig2}b) displays the spectrum of eigenvalues (on the left) and the resultant DOS (on the right) for 20\,meV  histogram slots near the valence- (VB) and conduction band (CB) edges for an intrinsic nanotube. The histogram unmistakably indicates the twofold degeneracy of all states. Such degeneracy, under the zone folding scheme, can be traced back to the K-point degeneracy present in the graphene band structure in reciprocal space.

Fig.~\ref{fig2}c), reveals the emergence of the $n=1$ Coulomb defect state which is split off from the valence band (VB) of the intrinsic system by 251\,meV. The aforementioned two-fold degeneracy of molecular orbitals is only slightly perturbed by the counterion, inducing an energy splitting in the $n=1$ state of no more than 2.5\,meV.

Figure \ref{fig2}d) illustrates the charge density contour for the $n=1$ defect state along the nanotube axis, having a full width at half maximum (FWHM) of 2.7\, nm. Figure \ref{fig2}e) offers an in-plane, false color depiction of the sign and amplitude of atomic $\pi$ orbitals that constitute the Coulomb defect state wavefunction. Less tightly bound, Rydberg-like states for $n=2,3,4,5$ — with the associated increased node count in the charge density contour — exhibit binding energies of 56, 27, 14, and 10\,meV, respectively. Intriguingly, these values align closely with a $n^{-2}$ scaling, analogous to higher lying states in hydrogenic systems (see supporting information for more details).

\begin{figure}[htbp]
	\centering
		\includegraphics[width=8.4 cm]{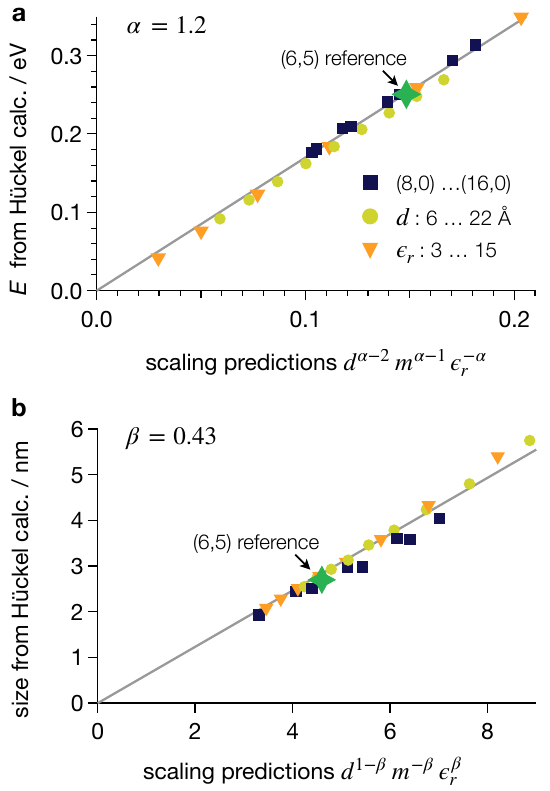}
        \caption{{\bf Comparing Modified H\"{u}ckel with Variational Calculations:} a) Modified H\"{u}ckel binding energies of the $n=1$ Coulomb defect. b) Size (FWHM) of charge density contours.}
        \label{fig3}
\end{figure}

To investigate the interplay between ion-surface distance, relative permittivity, and nanotube diameter on the binding energy and spatial extent of the most strongly bound defect state, we conducted calculations with each parameter being varied individually from the reference state while maintaining the others constant. In our assessment of tube diameter dependencies, (n,0) s-SWNT structures, bearing a similar count of C-atoms as the (6,5) s-SWNT, were employed. The results of the modified H\"{u}ckel calculations are compared in Figure \ref{fig3}a) with scaling predictions from a variational approach discussed further below. Notably, this reveals a characteristic scaling behaviour for the defect state binding energy with $E_b\propto d^{-0.8}\, m^{0.2}\,\epsilon_r^{-1.2}$, where $m$ is the effective hole mass. This underscores a nuanced relationship between these parameters and the defect state energy.

To uncover the observed scaling behaviors, we employed a variational approach to capture the essential features of the Coulomb defects. This model aims to describe the interaction between a $\pi$ electron (or hole) distributed over the $p_z$ orbitals in the three-dimensional nanotube structure and an exdohedral counterion carrying a charge $ez$. We simplify this by using a one-dimensional approximation in which a $\pi$ electron with effective band mass $m=m_{\rm eff}$ moves in the Coulomb potential $V(x)$ along the $x$-coordinate, aligned with the nanotube's longitudinal axis. The potential energy for the moving electron (or hole) in the Hamiltonian $\hat H$ is then determined by $e^2z / (4\pi\epsilon_0 \epsilon_r \sqrt{d^2+x^2})$, where $d$ is the counterion's distance from the $x$-axis. Inspired by the discussion of exciton scaling laws by Perebeinos et al.,\cite{Perebeinos2004} we employed the variational principle with a Gaussian trial function $\psi_\sigma\!(x)\propto \exp(-x^2/2\sigma^2)$ to elucidate the relevant scaling behaviors of Coulomb defects.

Accordingly, the total energy of the moving $\pi$ electron (or hole) can be expressed as a function of the variational parameter $\sigma$:
\begin{align}
    E_\sigma&=\langle \psi_\sigma|\hat H|\psi_\sigma\rangle/\langle \psi_\sigma|\psi_\sigma\rangle\nonumber \\
    &=\frac{\hbar^2}{4m\sigma^2}+\frac{e^2z}{4\pi\epsilon_0 \epsilon_r}\,\int_{-\infty}^{+\infty} dx\,\frac{|\psi_\sigma|^2}{\sqrt{d^2+x^2}}\nonumber \\
    &=\frac{\hbar^2}{4m\sigma^2}+\frac{e^2z}{\epsilon_r\sigma}\, f\!\left(\frac{d}{\sigma}\right)
    \label{eq2}
\end{align}
Reorganisation of the terms allows to rewrite this energy functional using a new two-dimensional function $h(mdz/\epsilon_r,d/\sigma)$ (see supporting information for details):
\begin{equation} 
    E_\sigma=\frac{1}{md^2}\,h\!\left(\frac{mdz}{\epsilon_r},\frac{d}{\sigma}\right)
    \label{eq3}
\end{equation}
$E_\sigma$ is now expressed as a function of two variables, $mdz/\epsilon_r$ and $d/\sigma$. Once minimized the resulting variational ground state energy, represented as $E \approx \min\limits_{\sigma} E_\sigma$, is approximated by a power law of $(mdz/\epsilon_r)$:
\begin{equation}
    E\approx A\, \frac{1}{m\,d^2}\,\left(\frac{m\,d\,|z|}{\epsilon_r}\right)^\alpha
    \label{eq4}
\end{equation}

Here, both $A$ and $\alpha$ are empirical parameters. For comparison of these scaling predictions with the H\"{u}ckel calculations in Fig.~\ref{fig3}, we calculated the effective band mass of holes across various tube types using curvature-corrected zone-folded tight-binding band structures~\cite{Ding2002,Hagen2003}.

Results from the modified H\"{u}ckel calculations depicted in Fig.~\ref{fig3}a) align remarkably well with the predicted scaling behavior when using $\alpha=1.2$. This value is somewhat smaller than $\alpha=2$ which is typical of the scaling behavior in hydrogenic systems. When expressing $d$ in nanometers and $m$ with respect to the electron mass $m_e$, the defect binding energy in eV is obtained from eq. \ref{eq4} using $A=1.70$. The legend in Fig.~\ref{fig3}a also indicates the range over which the parameters $\epsilon_r$, $d$, and the tube $(n,0)$ type were varied. 

Interestingly, similar scaling has been identified for the binding energies of excitons in s-SWNTs, with $\alpha_{\rm X}=1.40$~\cite{Perebeinos2004}. This commonality suggests that such scaling laws have broader applicability. The likely reason is that the physics of various types of two-particle Coulomb-bound systems are governed by Hamiltonians with similar distance, screening, charge and mass dependencies. Although excitons and Coulomb defects operate on slightly different yet related length scales, they are characterized by the typical distances assumed by the opposing charges. For excitons the critical length scale is the nanotube diameter, whereas for the Coulomb defects, it is the ion distance from the tube axis.

If we next look at the size of the defect states using its charge density contour, we propose that its full width at half maximum (FWHM) follows a related scaling law:  
\begin{equation}
    {\rm FWHM}\approx B\, d\,\left(\frac{m\,d\,| z|}{\epsilon_r}\right)^{-\beta}
    \label{eq5}
\end{equation}

Consistent with our earlier findings, the defect sizes computed via the H\"{u}ckel method align well with the scaling law expressed in Eq. \ref{eq5}, when $\beta$ is set to 0.43. This value is smaller than the value of $\beta=1$ characteristic of ideal hydrogenic problems. By expressing ion distance and effective mass in the same units as previously, we obtain the scaling of the $n=1$ defect states FWHM in nm using $B=0.616$.

These results offer valuable insights into experimental observations by highlighting the influence of environmental and structural factors on defect states. The character of counterions—including their charge number, radii, and solvation shells—along with the solvent's dielectric properties are essential in determining binding energies. In aqueous solutions, the high relative permittivity of $\epsilon_r\approx 80$ contrasts with considerably weaker screening found in organic solvents used for polymer-dispersed nanotubes. For example, in Poly(9,9-dioctylfluorene-co-benzothiadiazole) (PFO-BPy) dispersed s-SWNTs suspended in toluene and doped with $\rm Au(III)Cl_3$, the binding energy for defect states associated with adsorbed chlorine ions is estimated at roughly 100 meV \cite{Murrey2023}.

In support of these predictions, a study by Murrey et al. showed that sterically more distanced counterions, such as those associated with icosahedral dodecaborane (DDB) clusters, result in weaker binding of Coulomb defect states. In the study by Murrey et al. this was revealed by changes in the carrier transport properties of DDB-doped s-SWNTs, confirming the role of ion-tube separation in modulating Coulombic interactions~\cite{Murrey2023}.

\subsection{Interaction of Proximal Defects}
\begin{figure}[htbp]
	\centering
		\includegraphics[width=8.4 cm]{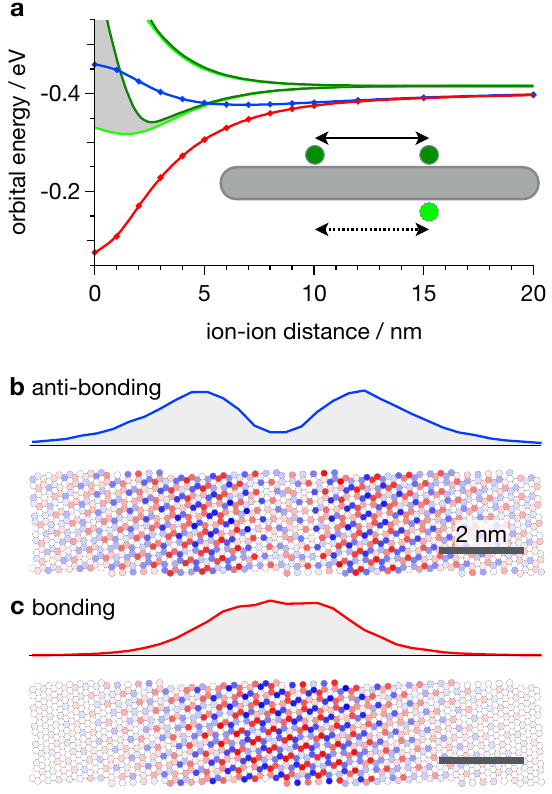}
        \caption{{\bf Proximal Defect States:} a) Bonding (red) and antibonding (blue) orbital hybridization with energy lines (green) for same-side and opposing counterions. b, c) Charge density and atomic amplitude for antibonding and bonding defect pairs at $s=2.7\,\rm nm$.}  
		\label{fig4}
\end{figure}

Now, we turn our attention to the influence of counterion proximity on the electronic structure of s-SWNTs, especially relevant at somewhat higher doping levels. As depicted in Figure \ref{fig4}, the orbital energies of the defect state undergo significant changes when the separation $s$ between two counterions on the nanotube surface is varied from 0 to 20 nm. For large counterion separations, the four $n=1$ defect states are nearly degenerate. In contrast, when the ions approach one another, these states split into two distinct pairs: a set exhibiting bonding and another showing anti-bonding character, evident by a state splitting reaching 380 meV for $s=0$ (see Fig.~\ref{fig4}a).

As the counterion spacing $s$ reaches zero, the bonding state becomes the $n=1$ ground state of the doubly charged ion - analogous to united atom correlations for the di-hydrogen ion $\rm H_2^+$. The binding energy of the bound state of 590 meV with respect to the VB edge, is consistent with the predicted $|z|^{\alpha}$ scaling with ion charge when $|z|=2$ and $\alpha=1.2$, as discussed above. Meanwhile, the antibonding state transitions into the $n=2$ defect state of the doubly charged ion.

At intermediate ion separations, the wavefunction of the overlapping defect states resemble the characteristics of bonding and antibonding molecular orbitals, akin to those in $\rm H_2^+$. When factoring in the screened counterion-repulsion, similarities also emerge in potential energy surfaces, again paralleling those known for $\rm H_2^+$. Fig.~\ref{fig4}a) displays two curves for both the bonding and antibonding orbitals, reflecting the Coulomb repulsion between counterions positioned either on opposite sides (light green curve) or the same side (dark green curve) of the nanotube. The charge density contours and false color representations of these molecular orbitals, when counterions are spaced 2.7 nm apart, are illustrated in Figs. \ref{fig4}b) and c). Notably, due to the nanotube's cylindrical form and chirality, the charge density contour of the antibonding molecular orbital does not feature a fully formed node.

These observations are particularly relevant for heavily doped semiconductors. In such cases, the electronic properties are dominated by the emergence of impurity bands, which arise from interactions between defect states. The origin of these bands parallels the interactions between defect pairs highlighted in our analysis.

\subsection{Conduction Band States}
While our discussion of Coulomb defects has primarily focused on modifications to valence band states, it is equally intriguing to examine the conduction band. 
\begin{figure}[htbp]
	\centering
		\includegraphics[width=8.4 cm]{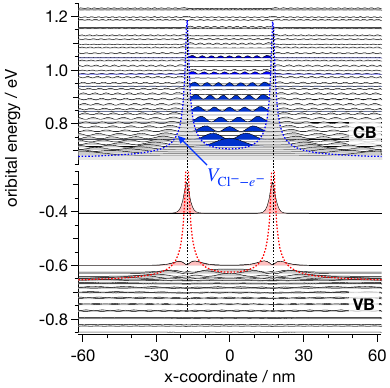}
        \caption{{\bf Wavefunctions Between Proximal Counterions:} The conduction band features new quantum well states induced by the repulsive electrostatic interaction with the counterions.}
        \label{fig5}
\end{figure}

Figure \ref{fig5} depicts charge density contours of H\"{u}ckel orbitals at energies ranging from -0.85 to +1.25 eV with respect to the intrinsic nanotube's band gap center. For the $s=35\,\rm nm$ counterion spacing considered, the computed orbital energy of the Coulomb defect states at the VB edge aligns closely with the energy of isolated defects.
\begin{figure}[htbp]
	\centering
		\includegraphics[width=8.4 cm]{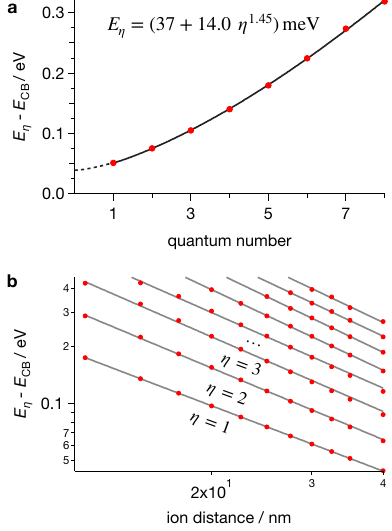}
        \caption{{\bf Scaling of Quantum Well States:} a) The energy of the lowest QW states scales with $\eta^{1.45}$ for a counterion spacing of 35\,nm. b) The dependence of these states on the separation of counterions, $s$, scales like $s^{-1.3}$ for the lowest QW states and like $s^{-1.4}$ for the higher QW states showcased here.}
        \label{fig6}
\end{figure} 

A closer inspection of conduction band (CB) states shown in Fig.~\ref{fig5}, reveals two distinct types of wavefunctions near the CB edge: quantum well (QW) like states (highlighted by blue fillings) that are localized between counterions and states with higher probability density outside this region. The blue dashed line in the conduction band represents the electrostatic interaction potential $V_{{\rm Cl}^--e^-}$ between an electron and the adjacent counterions. This potential's baseline aligns with the conduction band edge of the intrinsic nanotube.

Next we turn to the scaling behavior of the lowest QW states. In Fig.~\ref{fig6}a), the energies of the 8 lowest QW states are plotted against the quantum number $\eta$. Deviating somewhat from the infinite square potential well, the energy $E_\eta$ of the lowest state relative to the conduction band scales as $E_\eta=(37+14.0\,\eta^{1.45})\, \rm meV$ and not with $\eta^2$. Unsurprisingly, this stems from the discrepancy between the shape of the blue dashed interaction potential (seen in Fig.~\ref{fig5}) and a square potential well. Additional differences are the increasing transparency of the barriers as energies approach the peak of the potential energy curve at the countercharge locations.

Lastly, we evaluate the dependence of the QW states on counterion spacing, varying from 12 to 40 nm, as shown by Fig.~\ref{fig6}b). The binding energy of the $\eta=1$ ground state follows a $s^{-1.3}$ scaling while higher states are better described by $s^{-1.4}$ scaling. It is worth noting that valence band states do not exhibit such QW-like confinement, a consequence of the attractive hole-countercharge forces.

\subsection{Relevance for Exciton Transitions}
\begin{figure}[htbp]
	\centering
		\includegraphics[width=8.4 cm]{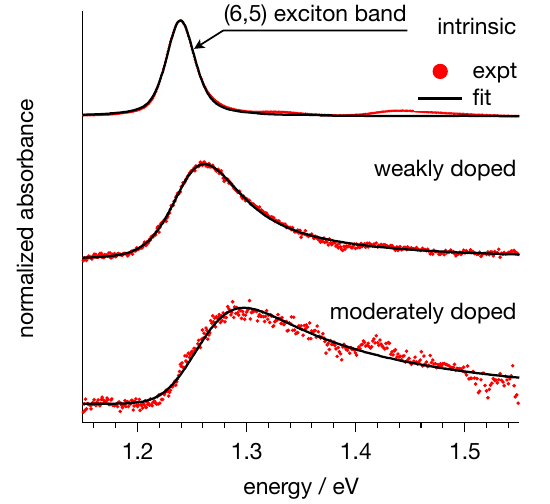}
        \caption{{\bf Experimentally Observed Shift and Broadening of the Exciton Band in (6,5) s-SWNTs:} The spectra of the weakly and moderately doped nanotubes are fitted by considering exciton confinement due to randomly placed counterions.}
		\label{fig7}
\end{figure}

The significance of having identified quantum well-like states in the conduction band becomes evident when considering their role in exciton transitions of redox-doped carbon nanotubes. These transitions involve superpositions of valence band (VB) and conduction band (CB) states \cite{Rohlfing2000}. As illustrated by the three normalized spectra of gold(III) chloride p-doped s-SWNT samples from reference \cite{Eckstein2019} in Fig.~\ref{fig7}, the near-infrared exciton bands of doped s-SWNTs undergo several changes. Specifically, these bands experience a blue-shift, broaden, and display increased asymmetry as doping levels rise.

The blue-shift of excitons in homogeneously doped systems is generally attributed to Pauli blocking \cite{Huard2000}, with band renormalization effects mitigating such shifts to some degree \cite{Spataru2005}. The effect of Pauli blocking eliminates single-particle states closest to the valence or conduction band edges from the superposition constituting the two-particle contributions to the exciton wavefunction $\Psi(\mathbf{r}_e, \mathbf{r}_h)$, as described, for example, by Rohlfing et al. \cite{Rohlfing2000}."
\begin{equation}
    \Psi\!(\mathbf{r}_e, \mathbf{r}_h) = \sum_{\mathbf{k}_e, \mathbf{k}_h} c_{\mathbf{k}_e, \mathbf{k}_h}\, \phi_{\mathbf{k}_e}\!(\mathbf{r}_e) \,\phi_{\mathbf{k}_h}\!(\mathbf{r}_h)
    \label{eq6}
\end{equation}
with $ \mathbf{r}_e $ and $ \mathbf{r}_h $ being the position vectors of the electron and hole, respectively, $ \phi_{\mathbf{k}_e}\!(\mathbf{r}_e) $ and $ \phi_{\mathbf{k}_h}\!(\mathbf{r}_h) $ representing the orbital components of wavefunctions of the electron and hole with wavevectors $ \mathbf{k}_e $ and $ \mathbf{k}_h $, respectively, and $ c_{\mathbf{k}_e, \mathbf{k}_h} $ denoting the coefficients for the probability amplitude of the electron and hole being in states $ \mathbf{k}_e $ and $ \mathbf{k}_h $, respectively.

For homogeneously doped (10,0) carbon nanotubes, it has been reported by Spataru and L\'{e}onard \cite{Spataru2005} that Pauli blocking near the valence or conduction band edges is responsible for both the blue-shift of exciton transitions and an increase in the electron-hole correlation length.

Departing from this established perspective, we present an alternative mechanism that is fundamentally rooted in confinement. Specifically, this mechanism attributes the observed shifts in exciton band frequencies to changes in the eigenenergies  $\epsilon_{\mathbf{k}_e}$ of the single-particle wavefunctions $\phi_{\mathbf{k}_e}$. This view provides a comprehensive explanation for the observed blue-shift in exciton transitions, as well as for their broadening and increasing asymmetry. Importantly, the proposed mechanism is intrinsically tied to inhomogeneities in the electronic structure of redox-doped s-SWNTs, which are explored in depth throughout this paper.

When fitting experimental exciton spectra, we thus employ a phenomenological approach to capture the effects of confinement on exciton band shape and position in inhomogeneously doped systems. Specifically, we model the spectrum by averaging over spectral features, $I_s$, that represent the optical response of quantum wells (QW) with different counterion spacings, $s$. Given the random distribution of counterion positions, we weight each spectral feature by its prevalence, $\rho(s)$, allowing us to derive the inhomogeneously broadened exciton band for the doped system.
\begin{equation}
    \bar I\!(\nu)\propto \int_0^\infty I_{s}\!(\nu)\rho\!(s)\,ds
    \label{eq7}
\end{equation}

Assuming a random distribution of counterions, we use a Poisson distribution to describe the frequency of QW widths, $\rho\!(s) \propto \exp[-s/\bar s]$, where $\bar s$ is the mean counterion separation. To model the spectra in Fig.~\ref{fig7}, we next approximate the energy shift of the exciton band in quantum wells of width $s$ via simple particle-in-a-box energy scaling. Using a Voigt fit to the reference spectrum $I_{\infty}$ from the intrinsic nanotube sample in Fig.~\ref{fig7}, we can then accurately model the spectra of weakly and moderately p-doped (6,5) s-SWNTs, thereby lending support to the proposed role of confinement for excited states in heterogeneously doped systems.

\section{Conclusions}

We have identified scaling laws that govern both the formation of Coulomb defects and the emergence of quantum well-like states in the conduction band of s-SWNTs. Our approach is based on a simple, computationally feasible model that here considers doping in the presence of adsorbed chlorine $\rm Cl^-$ ions. Inspired by previous work by Perebeinos et al. \cite{Perebeinos2004}, we have established scaling laws for both the binding energy $d^{\alpha-2}m^{\alpha-1}\epsilon_r^{-\alpha}|z|^\alpha$ and the dimensions $d^{1-\beta}m^{-\beta}\epsilon_r^{\beta}|z|^{-\beta}$ of these defect states. These laws provide potential guidelines for more tailored doping schemes.

The scaling laws presented here apply not only to Coulomb defects, but also to exciton transitions in s-SWNTs as described by Perebeinos et al.\cite{Perebeinos2004}, as well as to prototypical hydrogenic systems. This suggests potential applicability for related Coulomb defects and excitonic phenomena in other systems. These include atomically thin two-dimensional systems such as two-dimensional transition metal dichalcogenides and certain layered perovskites, as well as bulk semiconductors.

Our results also highlight the ability of Coulomb defects to create quantum well structures, leading to characteristic energy shifts among a subset of valence or conduction band states. The proposed mechanism, thought to account for spectral changes in low to moderately inhomogeneously doped semiconductors, may similarly have broader relevance. This extends to 1D and 2D atomically thin systems, as well as bulk semiconductor materials.

While our study provides valuable insights into the phenomena induced by doping in s-SWNTs, there is room for further exploration. Future work could benefit from the use of more refined quantum chemical calculations, improved representation of the solvent environment, and the development of comprehensive models to capture many-particle effects in the optical spectra of heterogeneously doped systems.

\begin{suppinfo}
Derivation of scaling laws from variational analysis and additional results on Rydberg like series of Coulomb defects.
\end{suppinfo}


\section{Acknowledgements}
We extend our sincere appreciation to R. Mitric for his careful review and invaluable insights, which have significantly contributed to the quality of this manuscript. K. E. and T.H. acknowledge financial support by the German National Science Foundation through the DFG GRK2112 and through grant HE 3355/4-1.

\bibliography{literature}

\providecommand{\latin}[1]{#1}
\makeatletter
\providecommand{\doi}
  {\begingroup\let\do\@makeother\dospecials
  \catcode`\{=1 \catcode`\}=2 \doi@aux}
\providecommand{\doi@aux}[1]{\endgroup\texttt{#1}}
\makeatother
\providecommand*\mcitethebibliography{\thebibliography}
\csname @ifundefined\endcsname{endmcitethebibliography}
  {\let\endmcitethebibliography\endthebibliography}{}
\begin{mcitethebibliography}{39}
\providecommand*\natexlab[1]{#1}
\providecommand*\mciteSetBstSublistMode[1]{}
\providecommand*\mciteSetBstMaxWidthForm[2]{}
\providecommand*\mciteBstWouldAddEndPuncttrue
  {\def\EndOfBibitem{\unskip.}}
\providecommand*\mciteBstWouldAddEndPunctfalse
  {\let\EndOfBibitem\relax}
\providecommand*\mciteSetBstMidEndSepPunct[3]{}
\providecommand*\mciteSetBstSublistLabelBeginEnd[3]{}
\providecommand*\EndOfBibitem{}
\mciteSetBstSublistMode{f}
\mciteSetBstMaxWidthForm{subitem}{(\alph{mcitesubitemcount})}
\mciteSetBstSublistLabelBeginEnd
  {\mcitemaxwidthsubitemform\space}
  {\relax}
  {\relax}

\bibitem[Bishop \latin{et~al.}(2020)Bishop, Hills, Srimani, Lau, Murphy,
  Fuller, Humes, Ratkovich, Nelson, and Shulaker]{Bishop2020}
Bishop,~M.~D.; Hills,~G.; Srimani,~T.; Lau,~C.; Murphy,~D.; Fuller,~S.;
  Humes,~J.; Ratkovich,~A.; Nelson,~M.; Shulaker,~M.~M. Fabrication of Carbon
  Nanotube Field-Effect Transistors in Commercial Silicon Manufacturing
  Facilities. \emph{Nat. Electron.} \textbf{2020}, \emph{3}, 492--501\relax
\mciteBstWouldAddEndPuncttrue
\mciteSetBstMidEndSepPunct{\mcitedefaultmidpunct}
{\mcitedefaultendpunct}{\mcitedefaultseppunct}\relax
\EndOfBibitem
\bibitem[Chen \latin{et~al.}(2016)Chen, Gao, Emaminejad, Kiriya, Ota, Nyein,
  Takei, and Javey]{Chen2016}
Chen,~K.; Gao,~W.; Emaminejad,~S.; Kiriya,~D.; Ota,~H.; Nyein,~H. Y.~Y.;
  Takei,~K.; Javey,~A. Printed Carbon Nanotube Electronics and Sensor Systems.
  \emph{Adv. Mater.} \textbf{2016}, \emph{28}, 4397--4414\relax
\mciteBstWouldAddEndPuncttrue
\mciteSetBstMidEndSepPunct{\mcitedefaultmidpunct}
{\mcitedefaultendpunct}{\mcitedefaultseppunct}\relax
\EndOfBibitem
\bibitem[Avouris \latin{et~al.}(2008)Avouris, Freitag, and
  Perebeinos]{Avouris2008}
Avouris,~P.; Freitag,~M.; Perebeinos,~V. Carbon-Nanotube Photonics and
  Optoelectronics. \emph{Nat. Photonics} \textbf{2008}, \emph{2},
  341--350\relax
\mciteBstWouldAddEndPuncttrue
\mciteSetBstMidEndSepPunct{\mcitedefaultmidpunct}
{\mcitedefaultendpunct}{\mcitedefaultseppunct}\relax
\EndOfBibitem
\bibitem[He \latin{et~al.}(2018)He, Htoon, Doorn, Pernice, Pyatkov, Krupke,
  Jeantet, Chassagneux, and Voisin]{He2018}
He,~X.; Htoon,~H.; Doorn,~S.~K.; Pernice,~W. H.~P.; Pyatkov,~F.; Krupke,~R.;
  Jeantet,~A.; Chassagneux,~Y.; Voisin,~C. Carbon Nanotubes as Emerging
  Quantum-Light Sources. \emph{Nat. Mater.} \textbf{2018}, \emph{17},
  663--670\relax
\mciteBstWouldAddEndPuncttrue
\mciteSetBstMidEndSepPunct{\mcitedefaultmidpunct}
{\mcitedefaultendpunct}{\mcitedefaultseppunct}\relax
\EndOfBibitem
\bibitem[Ishii \latin{et~al.}(2018)Ishii, He, Hartmann, Machiya, Htoon, Doorn,
  and Kato]{Ishii2018}
Ishii,~A.; He,~X.; Hartmann,~N.~F.; Machiya,~H.; Htoon,~H.; Doorn,~S.~K.;
  Kato,~Y.~K. Enhanced Single-Photon Emission from Carbon-Nanotube Dopant
  States Coupled to Silicon Microcavities. \emph{Nano Lett.} \textbf{2018},
  \emph{18}, 3873--3878\relax
\mciteBstWouldAddEndPuncttrue
\mciteSetBstMidEndSepPunct{\mcitedefaultmidpunct}
{\mcitedefaultendpunct}{\mcitedefaultseppunct}\relax
\EndOfBibitem
\bibitem[Ren \latin{et~al.}(2011)Ren, Bernardi, Lunt, Bulovic, Grossman, and
  Grade{\v{c}}ak]{Ren2011}
Ren,~S.; Bernardi,~M.; Lunt,~R.~R.; Bulovic,~V.; Grossman,~J.~C.;
  Grade{\v{c}}ak,~S. Toward Efficient Carbon Nanotube/P3HT Solar Cells: Active
  Layer Morphology, Electrical, and Optical Properties. \emph{Nano Lett.}
  \textbf{2011}, \emph{11}, 5316--5321\relax
\mciteBstWouldAddEndPuncttrue
\mciteSetBstMidEndSepPunct{\mcitedefaultmidpunct}
{\mcitedefaultendpunct}{\mcitedefaultseppunct}\relax
\EndOfBibitem
\bibitem[Kubie \latin{et~al.}(2018)Kubie, Watkins, Ihly, Wladkowski, Blackburn,
  Rice, and Parkinson]{Kubie2018}
Kubie,~L.; Watkins,~K.~J.; Ihly,~R.; Wladkowski,~H.~V.; Blackburn,~J.~L.;
  Rice,~W.~D.; Parkinson,~B.~A. Optically Generated Free-Carrier Collection
  from an All Single-Walled Carbon Nanotube Active Layer. \emph{J. Phys. Chem.
  Lett.} \textbf{2018}, \emph{9}, 4841--4847\relax
\mciteBstWouldAddEndPuncttrue
\mciteSetBstMidEndSepPunct{\mcitedefaultmidpunct}
{\mcitedefaultendpunct}{\mcitedefaultseppunct}\relax
\EndOfBibitem
\bibitem[Jain \latin{et~al.}(2012)Jain, Howden, Tvrdy, Shimizu, Hilmer,
  McNicholas, Gleason, and Strano]{Jain2012}
Jain,~R.~M.; Howden,~R.; Tvrdy,~K.; Shimizu,~S.; Hilmer,~A.~J.;
  McNicholas,~T.~P.; Gleason,~K.~K.; Strano,~M.~S. Polymer-Free Near-Infrared
  Photovoltaics with Single Chirality (6,5) Semiconducting Carbon Nanotube
  Active Layers. \emph{Adv. Mater.} \textbf{2012}, \emph{24}, 4436--4439\relax
\mciteBstWouldAddEndPuncttrue
\mciteSetBstMidEndSepPunct{\mcitedefaultmidpunct}
{\mcitedefaultendpunct}{\mcitedefaultseppunct}\relax
\EndOfBibitem
\bibitem[Pan \latin{et~al.}(2017)Pan, Li, and Choi]{Pan2017}
Pan,~J.; Li,~F.; Choi,~J.~H. Single-Walled Carbon Nanotubes as Optical Probes
  for Bio-Sensing and Imaging. \emph{J. Mater. Chem. B} \textbf{2017},
  \emph{5}, 6511--6522\relax
\mciteBstWouldAddEndPuncttrue
\mciteSetBstMidEndSepPunct{\mcitedefaultmidpunct}
{\mcitedefaultendpunct}{\mcitedefaultseppunct}\relax
\EndOfBibitem
\bibitem[Hu \latin{et~al.}(2003)Hu, Zhao, , Itkis, and Haddon]{Hu2003}
Hu,~H.; Zhao,~B.; ; Itkis,~M.~E.; Haddon,~R.~C. Nitric Acid Purification of
  Single-Walled Carbon Nanotubes. \emph{J. Phys. Chem. B} \textbf{2003},
  \emph{107}, 13838--13842\relax
\mciteBstWouldAddEndPuncttrue
\mciteSetBstMidEndSepPunct{\mcitedefaultmidpunct}
{\mcitedefaultendpunct}{\mcitedefaultseppunct}\relax
\EndOfBibitem
\bibitem[Nosho \latin{et~al.}(2007)Nosho, Ohno, Kishimoto, and
  Mizutani]{Nosho2007}
Nosho,~Y.; Ohno,~Y.; Kishimoto,~S.; Mizutani,~T. The Effects of Chemical Doping
  with F4TCNQ in Carbon Nanotube Field-Effect Transistors Studied by the
  Transmission-Line-Model Technique. \emph{Nanotechnology} \textbf{2007},
  \emph{18}, 415202\relax
\mciteBstWouldAddEndPuncttrue
\mciteSetBstMidEndSepPunct{\mcitedefaultmidpunct}
{\mcitedefaultendpunct}{\mcitedefaultseppunct}\relax
\EndOfBibitem
\bibitem[Kim \latin{et~al.}(2009)Kim, Jang, Kim, Park, Bae, Yu, Lee, Kim, Loc,
  Kim, Lee, Shin, Shin, Choi, and Lee]{Kim2009}
Kim,~S.~M.; Jang,~J.~H.; Kim,~K.~K.; Park,~H.~K.; Bae,~J.~J.; Yu,~W.~J.;
  Lee,~I.~H.; Kim,~G.; Loc,~D.~D.; Kim,~U.~J. \latin{et~al.}
  Reduction-Controlled Viologen in Bisolvent as an Environmentally Stable
  n-Type Dopant for Carbon Nanotubes. \emph{J. Am. Chem. Soc.} \textbf{2009},
  \emph{131}, 327--331\relax
\mciteBstWouldAddEndPuncttrue
\mciteSetBstMidEndSepPunct{\mcitedefaultmidpunct}
{\mcitedefaultendpunct}{\mcitedefaultseppunct}\relax
\EndOfBibitem
\bibitem[Chandra \latin{et~al.}(2010)Chandra, Afzali, Khare, El-Ashry, and
  Tulevski]{Chandra2010}
Chandra,~B.; Afzali,~A.; Khare,~N.; El-Ashry,~M.~M.; Tulevski,~G.~S. Stable
  Charge-Transfer Doping of Transparent Single-Walled Carbon Nanotube Films.
  \emph{Chem. Mater.} \textbf{2010}, \emph{22}, 5179--5183\relax
\mciteBstWouldAddEndPuncttrue
\mciteSetBstMidEndSepPunct{\mcitedefaultmidpunct}
{\mcitedefaultendpunct}{\mcitedefaultseppunct}\relax
\EndOfBibitem
\bibitem[Rao \latin{et~al.}(1997)Rao, Eklund, Bandow, Thess, and
  Smalley]{Rao1997}
Rao,~A.~M.; Eklund,~P.~C.; Bandow,~S.; Thess,~A.; Smalley,~R.~E. Evidence for
  Charge Transfer in Doped Carbon Nanotube Bundles from Raman Scattering.
  \emph{Nature} \textbf{1997}, \emph{388}, 257--259\relax
\mciteBstWouldAddEndPuncttrue
\mciteSetBstMidEndSepPunct{\mcitedefaultmidpunct}
{\mcitedefaultendpunct}{\mcitedefaultseppunct}\relax
\EndOfBibitem
\bibitem[Klinke \latin{et~al.}(2005)Klinke, Chen, Afzali, and
  Avouris]{Klinke2005}
Klinke,~C.; Chen,~J.; Afzali,~A.; Avouris,~P. Charge Transfer Induced Polarity
  Switching in Carbon Nanotube Transistors. \emph{Nano Lett.} \textbf{2005},
  \emph{5}, 555--558\relax
\mciteBstWouldAddEndPuncttrue
\mciteSetBstMidEndSepPunct{\mcitedefaultmidpunct}
{\mcitedefaultendpunct}{\mcitedefaultseppunct}\relax
\EndOfBibitem
\bibitem[Kim \latin{et~al.}(2008)Kim, Bae, Park, Kim, Geng, Park, Park, Shin,
  Shin, Yoon, Benayad, Choi, and Lee]{Kim2008}
Kim,~K.~K.; Bae,~J.~J.; Park,~H.~K.; Kim,~S.~M.; Geng,~H.-Z.; Park,~K.~A.;
  Park,~K.~A.; Shin,~H.~S.; Shin,~H.-J.; Yoon,~S.-M. \latin{et~al.}  Fermi
  Level Engineering of Single-Walled Carbon Nanotubes by AuCl\textsubscript{3}
  Doping. \emph{J. Am. Chem. Soc.} \textbf{2008}, \emph{130},
  12757--12761\relax
\mciteBstWouldAddEndPuncttrue
\mciteSetBstMidEndSepPunct{\mcitedefaultmidpunct}
{\mcitedefaultendpunct}{\mcitedefaultseppunct}\relax
\EndOfBibitem
\bibitem[Duong \latin{et~al.}(2010)Duong, Lee, Kim, Kong, Lee, and
  Lee]{Duong2010}
Duong,~D.~L.; Lee,~I.~H.; Kim,~K.~K.; Kong,~J.; Lee,~S.~M.; Lee,~Y.~H. Carbon
  Nanotube Doping Mechanism in a Salt Solution and Hygroscopic Effect: Density
  Functional Theory. \emph{ACS Nano} \textbf{2010}, \emph{4}, 5430--5436\relax
\mciteBstWouldAddEndPuncttrue
\mciteSetBstMidEndSepPunct{\mcitedefaultmidpunct}
{\mcitedefaultendpunct}{\mcitedefaultseppunct}\relax
\EndOfBibitem
\bibitem[Lee \latin{et~al.}(2010)Lee, Kim, Son, Yoon, Yao, Yu, Duong, Choi,
  Kim, Lee, and Lee]{Lee2010}
Lee,~I.~H.; Kim,~U.~J.; Son,~H.~B.; Yoon,~S.-M.; Yao,~F.; Yu,~W.~J.;
  Duong,~D.~L.; Choi,~J.-Y.; Kim,~J.~M.; Lee,~E.~H. \latin{et~al.}  Hygroscopic
  Effects on AuCl\textsubscript{3}-Doped Carbon Nanotubes. \emph{J. Phys. Chem.
  C} \textbf{2010}, \emph{114}, 11618--11622\relax
\mciteBstWouldAddEndPuncttrue
\mciteSetBstMidEndSepPunct{\mcitedefaultmidpunct}
{\mcitedefaultendpunct}{\mcitedefaultseppunct}\relax
\EndOfBibitem
\bibitem[Kim \latin{et~al.}(2011)Kim, Kim, Jo, Park, Chae, Duong, Yang, Kong,
  and Lee]{Kim2011}
Kim,~S.~M.; Kim,~K.~K.; Jo,~Y.~W.; Park,~M.~H.; Chae,~S.~J.; Duong,~D.~L.;
  Yang,~C.-W.; Kong,~J.; Lee,~Y.~H. Role of Anions in the
  AuCl\textsubscript{3}-Doping of Carbon Nanotubes. \emph{ACS Nano}
  \textbf{2011}, \emph{5}, 1236--1242\relax
\mciteBstWouldAddEndPuncttrue
\mciteSetBstMidEndSepPunct{\mcitedefaultmidpunct}
{\mcitedefaultendpunct}{\mcitedefaultseppunct}\relax
\EndOfBibitem
\bibitem[Murat \latin{et~al.}(2014)Murat, Rungger, Jin, Jin, Sanvito, and
  Schwingenschlögl]{Murat2014}
Murat,~A.; Rungger,~I.; Jin,~C.; Jin,~C.; Sanvito,~S.; Schwingenschlögl,~U.
  Origin of the p-Type Character of AuCl3 Functionalized Carbon Nanotubes.
  \emph{J. Phys. Chem. C} \textbf{2014}, \emph{118}, 3319--3323\relax
\mciteBstWouldAddEndPuncttrue
\mciteSetBstMidEndSepPunct{\mcitedefaultmidpunct}
{\mcitedefaultendpunct}{\mcitedefaultseppunct}\relax
\EndOfBibitem
\bibitem[Hertel(2019)]{Hertel2019}
Hertel,~T. In \emph{Optical Properties of Carbon Nanotubes}; Weisman,~R.~B.,
  Kono,~J., Eds.; World Scientific Series on Carbon Nanoscience, Handbook of
  Carbon Nanomaterials; {World Scientific}: New Jersey, 2019; Vol.~10; pp
  191--236\relax
\mciteBstWouldAddEndPuncttrue
\mciteSetBstMidEndSepPunct{\mcitedefaultmidpunct}
{\mcitedefaultendpunct}{\mcitedefaultseppunct}\relax
\EndOfBibitem
\bibitem[Mouri \latin{et~al.}(2013)Mouri, Miyauchi, Iwamura, and
  Matsuda]{Mouri2013}
Mouri,~S.; Miyauchi,~Y.; Iwamura,~M.; Matsuda,~K. Temperature Dependence of
  Photoluminescence Spectra in Hole-Doped Single-Walled Carbon Nanotubes:
  Implications of Trion Localization. \emph{Phys. Rev. B} \textbf{2013},
  \emph{87}, 045408\relax
\mciteBstWouldAddEndPuncttrue
\mciteSetBstMidEndSepPunct{\mcitedefaultmidpunct}
{\mcitedefaultendpunct}{\mcitedefaultseppunct}\relax
\EndOfBibitem
\bibitem[Eckstein \latin{et~al.}(2017)Eckstein, Hartleb, Achsnich,
  Sch{\"o}ppler, and Hertel]{Eckstein2017}
Eckstein,~K.~H.; Hartleb,~H.; Achsnich,~M.~M.; Sch{\"o}ppler,~F.; Hertel,~T.
  Localized Charges Control Exciton Energetics and Energy Dissipation in Doped
  Carbon Nanotubes. \emph{ACS Nano} \textbf{2017}, \emph{10},
  10401--10408\relax
\mciteBstWouldAddEndPuncttrue
\mciteSetBstMidEndSepPunct{\mcitedefaultmidpunct}
{\mcitedefaultendpunct}{\mcitedefaultseppunct}\relax
\EndOfBibitem
\bibitem[Eckstein \latin{et~al.}(2019)Eckstein, Oberndorfer, Achsnich,
  Sch{\"o}ppler, and Hertel]{Eckstein2019}
Eckstein,~K.~H.; Oberndorfer,~F.; Achsnich,~M.~M.; Sch{\"o}ppler,~F.;
  Hertel,~T. Quantifying Doping Levels in Carbon Nanotubes by Optical
  Spectroscopy. \emph{J. Phys. Chem. C} \textbf{2019}, \emph{123},
  30001--30006\relax
\mciteBstWouldAddEndPuncttrue
\mciteSetBstMidEndSepPunct{\mcitedefaultmidpunct}
{\mcitedefaultendpunct}{\mcitedefaultseppunct}\relax
\EndOfBibitem
\bibitem[Eckstein \latin{et~al.}(2021)Eckstein, Hirsch, Martel, and
  Hertel]{Eckstein2021}
Eckstein,~K.~H.; Hirsch,~F.; Martel,~R.; Hertel,~T. Infrared Study of Charge
  Carrier Confinement in Doped (6,5) Carbon Nanotubes. \emph{J. Phys. Chem. C}
  \textbf{2021}, \emph{125}, 5700--5707\relax
\mciteBstWouldAddEndPuncttrue
\mciteSetBstMidEndSepPunct{\mcitedefaultmidpunct}
{\mcitedefaultendpunct}{\mcitedefaultseppunct}\relax
\EndOfBibitem
\bibitem[Perebeinos \latin{et~al.}(2004)Perebeinos, Tersoff, and
  Avouris]{Perebeinos2004}
Perebeinos,~V.; Tersoff,~J.; Avouris,~P. Scaling of Excitons in Carbon
  Nanotubes. \emph{Phys. Rev. Lett.} \textbf{2004}, \emph{92}, 257402\relax
\mciteBstWouldAddEndPuncttrue
\mciteSetBstMidEndSepPunct{\mcitedefaultmidpunct}
{\mcitedefaultendpunct}{\mcitedefaultseppunct}\relax
\EndOfBibitem
\bibitem[Ding \latin{et~al.}(2002)Ding, Yan, and Cao]{Ding2002}
Ding,~J.~W.; Yan,~X.~H.; Cao,~J. Analytical Relation of Band Gaps to Both
  Chirality and Diameter of Single-Wall Carbon Nanotubes. \emph{Phys. Rev. B}
  \textbf{2002}, \emph{66}, 073401\relax
\mciteBstWouldAddEndPuncttrue
\mciteSetBstMidEndSepPunct{\mcitedefaultmidpunct}
{\mcitedefaultendpunct}{\mcitedefaultseppunct}\relax
\EndOfBibitem
\bibitem[Hagen and Hertel(2003)Hagen, and Hertel]{Hagen2003}
Hagen,~A.; Hertel,~T. Quantitative Analysis of Optical Spectra from Individual
  Single-Wall Carbon Nanotubes. \emph{Nano Lett.} \textbf{2003}, \emph{3},
  383--388\relax
\mciteBstWouldAddEndPuncttrue
\mciteSetBstMidEndSepPunct{\mcitedefaultmidpunct}
{\mcitedefaultendpunct}{\mcitedefaultseppunct}\relax
\EndOfBibitem
\bibitem[Pedersen(2004)]{Pedersen2004}
Pedersen,~T.~G. Exciton Effects in Carbon Nanotubes. \emph{Carbon}
  \textbf{2004}, \emph{42}, 1007--1010\relax
\mciteBstWouldAddEndPuncttrue
\mciteSetBstMidEndSepPunct{\mcitedefaultmidpunct}
{\mcitedefaultendpunct}{\mcitedefaultseppunct}\relax
\EndOfBibitem
\bibitem[H\"{u}ckel(1931)]{Huckel1931}
H\"{u}ckel,~E. Quantentheoretische Beitr\"{a}ge zum Benzolproblem. I. Die
  Elektronenkonfiguration des Benzols und werwandter Verbindungen. \emph{Z.
  Phys. Chem.} \textbf{1931}, \emph{70}, 204--286\relax
\mciteBstWouldAddEndPuncttrue
\mciteSetBstMidEndSepPunct{\mcitedefaultmidpunct}
{\mcitedefaultendpunct}{\mcitedefaultseppunct}\relax
\EndOfBibitem
\bibitem[Hoffmann(1963)]{Hoffmann1963}
Hoffmann,~R. An Extended Hückel Theory. I. Hydrocarbons. \emph{J. Chem. Phys.}
  \textbf{1963}, \emph{39}, 1397--1412\relax
\mciteBstWouldAddEndPuncttrue
\mciteSetBstMidEndSepPunct{\mcitedefaultmidpunct}
{\mcitedefaultendpunct}{\mcitedefaultseppunct}\relax
\EndOfBibitem
\bibitem[Saito \latin{et~al.}(1992)Saito, Fujita, Dresselhaus, and
  Dresselhaus]{Saito1992}
Saito,~R.; Fujita,~M.; Dresselhaus,~G.; Dresselhaus,~M.~S. Electronic Structure
  of Graphene Tubules Based on C\textsubscript{60}. \emph{Phys. Rev. B}
  \textbf{1992}, \emph{46}, 1804--1811\relax
\mciteBstWouldAddEndPuncttrue
\mciteSetBstMidEndSepPunct{\mcitedefaultmidpunct}
{\mcitedefaultendpunct}{\mcitedefaultseppunct}\relax
\EndOfBibitem
\bibitem[Hamada \latin{et~al.}(1992)Hamada, Sawada, and Oshiyama]{Hamada1992}
Hamada,~N.; Sawada,~S.; Oshiyama,~A. New One-Dimensional Conductors: Graphitic
  Microtubules. \emph{Phys. Rev. Lett.} \textbf{1992}, \emph{68},
  1579--1581\relax
\mciteBstWouldAddEndPuncttrue
\mciteSetBstMidEndSepPunct{\mcitedefaultmidpunct}
{\mcitedefaultendpunct}{\mcitedefaultseppunct}\relax
\EndOfBibitem
\bibitem[Wallace(1947)]{Wallace1947}
Wallace,~P.~R. The Band Theory of Graphite. \emph{Phys. Rev.} \textbf{1947},
  \emph{71}, 622--634\relax
\mciteBstWouldAddEndPuncttrue
\mciteSetBstMidEndSepPunct{\mcitedefaultmidpunct}
{\mcitedefaultendpunct}{\mcitedefaultseppunct}\relax
\EndOfBibitem
\bibitem[Murrey \latin{et~al.}(2023)Murrey, Aubry, Ruiz, Thurman, Eckstein,
  Doud, Stauber, Spokoyny, Schwartz, Hertel, Blackburn, and
  Ferguson]{Murrey2023}
Murrey,~T.~L.; Aubry,~T.~J.; Ruiz,~O.~L.; Thurman,~K.~A.; Eckstein,~K.~H.;
  Doud,~E.~A.; Stauber,~J.~M.; Spokoyny,~A.~M.; Schwartz,~B.~J.; Hertel,~T.
  \latin{et~al.}  Tuning Counterion Chemistry to Reduce Carrier Localization in
  Doped Semiconducting Carbon Nanotube Networks. \emph{Cell Rep. Phys. Sci.}
  \textbf{2023}, \emph{4}, 101407\relax
\mciteBstWouldAddEndPuncttrue
\mciteSetBstMidEndSepPunct{\mcitedefaultmidpunct}
{\mcitedefaultendpunct}{\mcitedefaultseppunct}\relax
\EndOfBibitem
\bibitem[Rohlfing and Louie(2000)Rohlfing, and Louie]{Rohlfing2000}
Rohlfing,~M.; Louie,~S.~G. Electron-Hole Excitations and Optical Spectra from
  First Principles. \emph{Phys. Rev. B} \textbf{2000}, \emph{62},
  4927--4944\relax
\mciteBstWouldAddEndPuncttrue
\mciteSetBstMidEndSepPunct{\mcitedefaultmidpunct}
{\mcitedefaultendpunct}{\mcitedefaultseppunct}\relax
\EndOfBibitem
\bibitem[Huard \latin{et~al.}(2000)Huard, Cox, Saminadayar, Arnoult, Arnoult,
  and Tatarenko]{Huard2000}
Huard,~V.; Cox,~R.; Saminadayar,~K.; Arnoult,~A.; Arnoult,~A.; Tatarenko,~S.
  Bound States in Optical Absorption of Semiconductor Quantum Wells Containing
  a Two-Dimensional Electron Gas. \emph{Phys. Rev. Lett.} \textbf{2000},
  \emph{84}, 187--190\relax
\mciteBstWouldAddEndPuncttrue
\mciteSetBstMidEndSepPunct{\mcitedefaultmidpunct}
{\mcitedefaultendpunct}{\mcitedefaultseppunct}\relax
\EndOfBibitem
\bibitem[Spataru \latin{et~al.}(2005)Spataru, Ismail-Beigi, Capaz, and
  Louie]{Spataru2005}
Spataru,~C.~D.; Ismail-Beigi,~S.; Capaz,~R.~B.; Louie,~S.~G. Theory and Ab
  Initio Calculation of Radiative Lifetime of Excitons in Semiconducting Carbon
  Nanotubes. \emph{Phys. Rev. Lett.} \textbf{2005}, \emph{95}, 247402\relax
\mciteBstWouldAddEndPuncttrue
\mciteSetBstMidEndSepPunct{\mcitedefaultmidpunct}
{\mcitedefaultendpunct}{\mcitedefaultseppunct}\relax
\EndOfBibitem
\end{mcitethebibliography}

\end{document}